\def\be{\begin{equation}}
\def\ee{\end{equation}}
\def\ber{\begin{eqnarray}}
\def\eer{\end{eqnarray}}
\def\bers{\begin{eqnarray*}}
\def\eers{\end{eqnarray*}}

\def\PRL{{ Phys. Rev. Lett.}\ }

\documentclass[aps,prb,twocolumn,showpacs,groupedaddressamsmath,amssymb]{revtex4-1}
\usepackage{dcolumn}   
\usepackage{amsmath}
\usepackage{graphicx}    
\usepackage{subfigure}
\usepackage{comment}
\usepackage{color}
\usepackage{ifthen}
\newboolean{includefigs}
\setboolean{includefigs}{true}      
\newboolean{includetext}
\setboolean{includetext}{true}     
\newcommand{\condcomment}[2]{\ifthenelse{#1}{#2}{}}
\begin{document}

\title{Half-metallic, Co-based quaternary Heuslers for spintronics: defect- and pressure-induced transitions and properties}
\author{ Enamullah,$^{1}$ D. D. Johnson,$^{2,3}$ K. G. Suresh,$^{1}$ and Aftab Alam$^{1}$}
\email{{\it aftab@phy.iitb.ac.in}, enamullah@phy.iitb.ac.in, ddj@ameslab.gov}
\affiliation{$^{1}$Department of Physics, Indian Institute of Technology, Bombay, Powai, Mumbai 400 076, India}
\affiliation{$^{2}$Division of Materials Science $\&$ Engineering, Ames Laboratory, Ames, Iowa 50011, USA}
\affiliation{$^{3}$Department of Materials Science $\&$ Engineering, Iowa State University, Ames, Iowa 50011, USA}

\begin{abstract}
{
Heusler compounds offer potential as spintronic devices due to their spin-polarization and half-metallicity properties, where electron spin-majority (minority) manifold exhibits states (band gap) at the electronic chemical potential, yielding full spin-polarization in a single manifold.  Yet, Heuslers often exhibit intrinsic disorder that degrades its half-metallicity and spin-polarization.  Using density-functional theory, we analyze the electronic and magnetic properties of  equiatomic Heusler ($L$2$_{1}$) CoMnCrAl and CoFeCrGe alloys for effects of hydrostatic pressure and intrinsic disorder (thermal antisites, binary swaps, and vacancies). 
Under pressure, CoMnCrAl undergoes a metallic transition, while half-metallicity in CoFeCrGe is retained for a limited range. 
Antisite disorder between Co-Al pairs in CoMnCrAl  and Co-Ge pairs in CoFeCrGe is energetically the most favored, and retain half-metallic character in Co-excess samples. However, Co-deficient samples undergo a transition from half-metallic to  metallic,  with  a discontinuity in the saturation magnetization. 
For binary swaps, configurations that compete with the ground state are identified and show no loss of half-metallicity; however, the minority-spin bandgap and magnetic moments vary depending on the atoms swapped. For single binary swaps, there is a significant energy cost in CoMnCrAl but with no loss of half metallicity. Although a few configurations in CoFeCrGe energetically compete with the ground statei, however the minority-spin bandgap and magnetic moments vary depending on the atoms swapped. These informations should help in controlling these potential spintronic materials.
}
\end{abstract}

\date{\today}
\pacs{31.15.A-, 85.75.-d, 75.50.Cc, 61.72.-y}
\maketitle
\section{Introduction}
{\par}   Disorder is an inherent property of any real material. Physical properties of functional materials e.g. conductivity, magnetization, are strongly influenced by impurities and point defects.  For spintronic based materials, it becomes even more important because all the phenomena are related to spin degrees of freedom (magnetization). The precise control of impurity species and concentrations in semiconductors underlies the fabrication of virtually all electronic and magneto electronic devices.
In terms of electron density $n({E}_\text{F})$ at the Fermi energy, $E_\text{F}$, half-metallicity arises due to a finite $n({E}_\text{F})$ in the majority-spin manifold and a bandgap in the minority-spin manifold.  Ideally, then, the spin polarization should be 100\% in half-metallic compounds. Experimentally  it is found to be 50-70\% because of chemical disorder.\cite{Raph1,Rav1,Raph2}  Thus, the half-metallic property plays a decisive role for magneto-electronics and spin-transport phenomena. 

{\par}  Half-metallicity in Heusler alloys (HAs), discovered by Groot {\it et al.},\cite{Groot} is formed by transition metals with $p$-block elements. Half-metallic and ferromagnetic properties are widely found in perovskite compounds,{\cite{Zhu,Kob}} metallic oxides,{\cite{Dho,Soe}} HAs,{\cite{Nou,Wan}} and magnetic semiconductors.{\cite{Kro,Noo}}
Amongst all systems, HAs are most favorable because of their high Curie temperatures and spin polarization along with the structural compatibility to the conventional wide-gap semiconductors.\cite{Far,Has1,Has2,Has3}

{\par}  The conventional HAs have 2:1:1 stoichiometry, i.e.,  $X_{2}YZ$ (ternary), with ordered $L$2$_{1}$ structure ($Fm\bar{3}m$, space group \#225), where $X$,$Y$ are  $d$-band metals and $Z$ is a non-magnetic $p$-block element. A 1:1:1:1 stoichiometric structure arises when one $X$ is replaced by a more or less electronegative, transition metal element, forming a $Y$-type structure ($F\bar{4}3m$ space group, \#216)
-- or equiatomic quaternary HAs.{\cite{Ozd,Ali,Dai,Sin,Ali1,Gal,Luo,Gal1,Mei}}

{\par} Neutron diffraction experiment on Co$_{2}$MnSi show 14\% of Mn sites are occupied by Co and 5-7\% of Co sites by Mn.{\cite{Raph2}} Similar results were observed by EXAFS.{\cite{Rav2}}  Distribution of transition metals ($X^{1}$, $X^{2}$ and $Y$) among each other induces disorder  and yields a $DO_{3}$ structure.{\cite{TG}} 
 When $X^{1}$=$X^{2}$ and $Y$=$Z$, $B$2 is formed,{\cite{TG}} whereas $A$2 forms when $X^{1}$=$X^{2}$=$Y$=$Z$ at all sites.{\cite{TG}}  Any antisite  disorder reduces  spin polarization in conventional HAs.
For example, Co antisites in  (i) Co$_{2}$MnGe cause loss of half-metallicity,{\cite{Pico1}} and (ii) in Co$_{2}$MnSi{\cite{Pico2}}  reduce the minority band gap. 
Another type of intrinsic defect is swapping of two atoms from their preferred Wyckoff site, which lowers the half-metallic property of HAs and reduces the minority-spin gap states, as happens when Co-Mn and Mn-Si swap in Co$_{2}$MnSi. Vacancies are also ubiquitously found in HAs,\cite{Hamad11} and often degrade their properties.    Despite the extensive studies on various intrinsic defects in ternary HAs, similar studies are missing for quaternary HAs. It is thus imperative to identify and precisely control point defects in such functional materials, and is one of the main focus of the present article.

{\par} A signature of half-metallicity in CoMnCrAl and CoFeCrGe was given elsewhere.\cite{enam} 
   Evidence of intrinsic disorder was suspected to destroy half-metallicity due to defect-induced minority gap states.
Here, in both CoMnCrAl and CoFeCrGe hosts, we systematically investigate the effects of various defects (antisites, binary swaps, and vacancies) and hydrostatic pressure (volume reduction) on the formation energy and defect-induced electronic and magnetic properties.  We found that CoMnCrAl (CoFeCrGe) is extremely sensitive to pressure and undergoes a transition from half-metallic to metallic state by $\sim 3\%$ ($\sim 7\%$) of lattice constant reduction. 
Antisite disorder between Co-Al pairs in CoMnCrAl and Co-Ge pairs in CoFeCrGe are the most favorable, and expected in real samples. This was suggested experimentally in CoMnCrAl,\cite{enam}  where antisite mixing of Al with other transition metals was suspected. Such defects beyond a certain concentration change gross properties,  e.g., loss of half-metallicity at $\sim 7.4\%$ Al-excess in CrCo$_{1-x}$Al$_{1+x}$Mn and $\sim 3.7\%$ Ge-excess in CrCo$_{1-x}$Ge$_{1+x}$Fe.

\begin{figure}[t]
\centering
\includegraphics[width=0.25\textwidth]{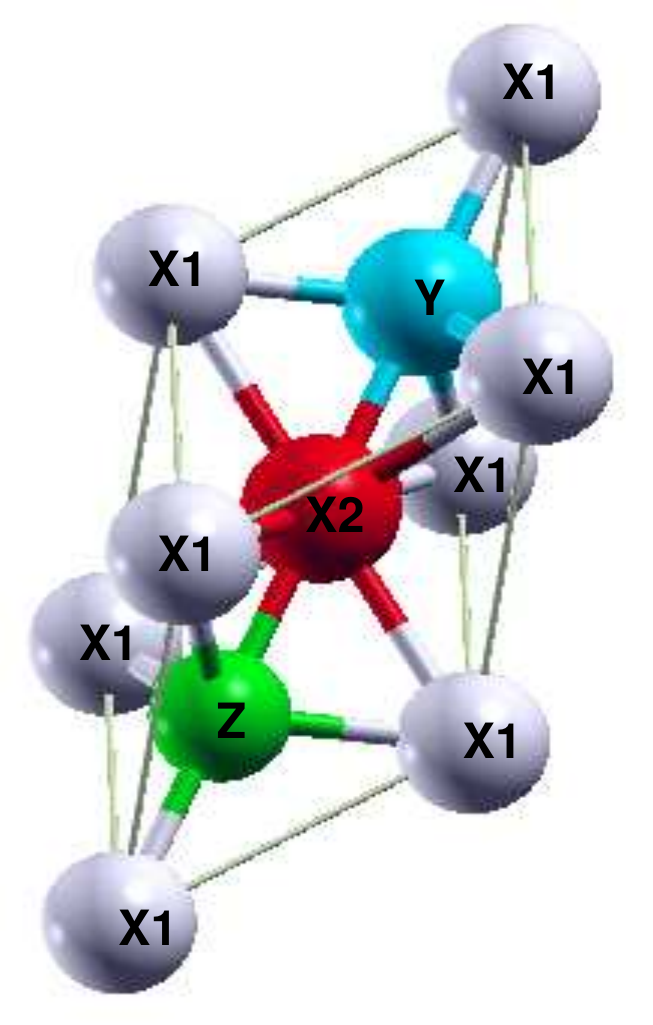}
\caption{ A $4$-atom fcc primitive cell of $X_{1}X_{2}YZ$ Heusler. }
\label{x1x2yz-333}
\end{figure}

\section{Computational Details}
{\par}
{\it{Ab}} {\it{initio}} simulations are performed by using a spin-polarized density functional theory (DFT) implemented within
Vienna ${\it{ab}}$ ${\it{initio}}$ simulation package (VASP)\cite{VASP} with a projected augmented-wave basis.\cite{PAW} We adopt the idea of generalized gradient approximation (GGA) in the scheme of Perdew-Burke-Ernzerhof (PBE) for the electronic exchange-correlation functional.
We used a plane wave cut-off of $368$ eV with the convergence to 0.1 meV/cell (10 kBar) for energy (stress). 
All calculations are fully relaxed. 
 Cubic lattice symmetry is preserved for most defects,  except for few where the unit cell angles  ($\alpha, \beta, \gamma$) narrowly deviate to  $89.7^0 - 89.9^{0}$ (compared to $90^{0}$ for cubic), e.g., antisites between Co-Ge and Fe-Ge in CoFeCrGe and Co-Al and Mn-Al in CoMnCrAl.

{\par} The $X^1$$X^2$$YZ$ CoMnCrAl and CoFeCrGe  have LiMgPdSn prototype ($Y$-type) cubic structure.
Site-preference energies suggest that the most stable structure are the ones with $X^{1}$ at $4c$, $X^{2}$ at $4a$, $Y$ at $4b$, and $Z$ at $4d$ Wyckoff sites.{\cite{enam}} 
 Because the amount of intrinsic defects in real systems are small, we simulate these defects in a 3 $\times$ 3 $\times$ 3 supercell, formed from a 4-atom fcc cell (see Fig.~\ref{x1x2yz-333}) of the most stable configurations.\cite{enam}  
 This supercell contains a total of $108$ atoms with $27$ atoms of each kind.  Brillouin zone integrations are performed using $24^{3}$ ($8^{3}$) k-mesh for 4-atom (108-atom) cells. Antisite defects amongst all pairs ($6$ in a quaternary), single and double vacancies,  and all combinations of binary swaps between different atoms have been investigated.  

{\par}  Relative stability of these defected cells were assessed by formation energies ($\Delta E_f$) that is referenced to the perfectly ordered endpoints, i.e., $X^1_2$$YZ$  and $X^2_2$$YZ$).    For a given antisite binary disorder $x$ in a $\text{A}_{1-x}\text{B}_{1+x} \text{C}\text{D}$  Heusler, $\Delta E_f$ is defined as :
\begin{eqnarray}
\Delta E_f &=& E[\text{A}_{1-x}\text{B}_{1+x} \text{C}\text{D}] -
\frac{1}{2} \left[\rule{0mm}{3.5mm}(1-x)\ E(\text{A}_2\text{C}\text{D}) \right. \nonumber\\
&&\left. +\ (1+x)\ E(\text{B}_2\text{C}\text{D}) \rule{0mm}{3.5mm}\right]. 
\end{eqnarray}
We have also checked the mechanical stability of the parent compounds by satisfying the Born-Huang criteria.\cite{BHC} This requires computing the elastic constants by performing a lattice dynamics calculation. Such calculations are computationally more expensive, and need higher accuracy. As such, we have used an energy cut-off of 500 eV, total energy convergence  of 10$^{-5}$ eV along with 8$^3$ k-mesh for BZ integration.

\section{Results and Discussion}
 We present the effect of hydrostatic pressure and point defects (antisite, swap and vacancy) for the two representative HAs, CoMnCrAl and CoFeCrGe. Both systems are of interest because a few preliminary experimental results\cite{enam} exist and they provide a platform for verifying our theoretical predictions. CoFeCrGe has a high Curie temperature ($T_\text{C} \sim  866$ K), and hence useful for high-temperature applications.
\subsection{CoMnCrAl}
\subsubsection{Pressure effect}
 Our calculated equilibrium lattice parameter $a_0$ of CoMnCrAl is $5.70 $ \r{A}, while the measured value is $5.77$ \r{A} at 300 K. We found half-metallic character at both of these lattice constants with corresponding minority-spin band gap ($\Delta E_\text{g}$)$_\downarrow = 0.24$ eV and $0.33$ eV, respectively.

\begin{figure}[t]
\centering
\includegraphics[scale=1.0]{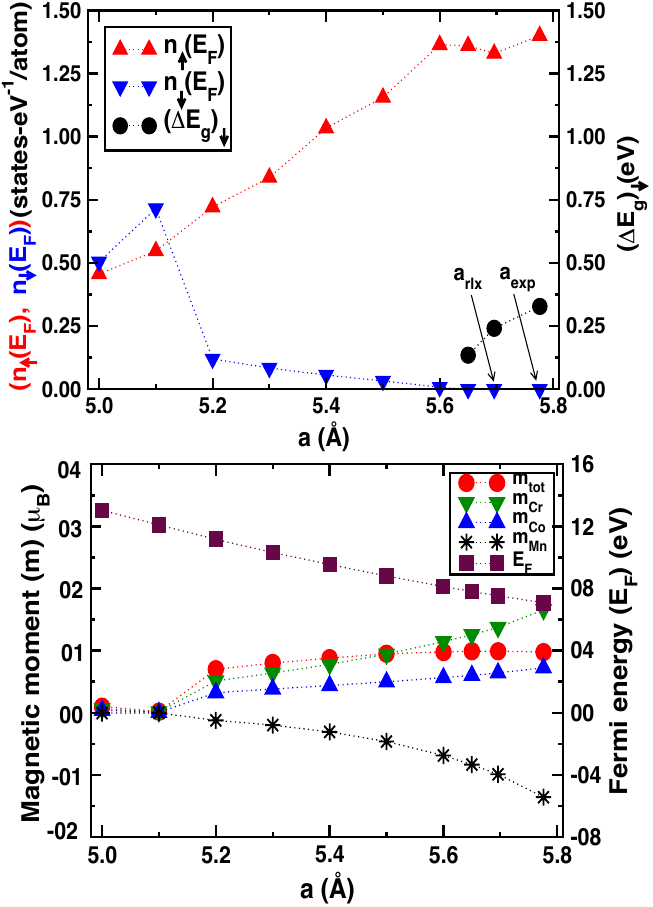}
\caption{For CoMnCrAl, pressure effect on  (top) density of states at $E_{\text{F}}$ for majority-spin n$_{\uparrow}(E_\text{F})$ (triangle up), minority-spin n$_{\downarrow}(E_\text{F})$ (down triangle), and minority-spin gap ($\Delta E_\text{g}$)$_{\downarrow}$ (circle); (bottom) total, atom-projected magnetic moments ($m$) and $E_{\text{F}}$. At 5.62 $\text{\r{A}}$, the transition from half-metallic to metallic occurs, where equilibrium lattice constant is $a_{\text{rlx}}=5.70$ \r{A}  ($a_{\text{exp}}=5.78$ \r{A}). }
\label{press_CMCA}
\end{figure}

{\par} To investigate the effect of hydrostatic pressure, we calculated the electronic structure at decreased $a$. Figure \ref{press_CMCA} shows the density of states (DoS) at $E_{\text F}$ for majority and minority spin channel as well as the band-gap in minority channel versus $a$ (or pressure). The system retains its half-metallicity in the vicinity of experimental $a_0$.
Figure \ref{press_CMCA} also shows the variation of total and atom-projected magnetization and $E_{\text F}$ vs. $a$. Notice that the system retains half-metallicity down to $\sim 5.62$ \r{A}, below which the minority spin exhibit a finite DoS at $E_{\text F}$, with a loss of band gap. This causes a transition to metallic behavior. Note that CoMnCrAl is quite sensitive to pressure because even a $2-3 \%$ reduction in $a$  the system transforms from half-metallic to metallic. 

{\par} Mn is antiferromagnetically aligned compared to Co and Cr (Fig. \ref{press_CMCA}). Up to $5.2$ \r{A}, the total moment does not vary much and follows the Slater-Pauling (SP) rule.\cite{Slat1,Paul1} Down to $5.2$ \r{A}, the moment collapses and the alloy becomes nonmagnetic. Such a huge pressure may not be achievable in experiments.
\begin{figure}[t]
\centering
\includegraphics[scale=0.48]{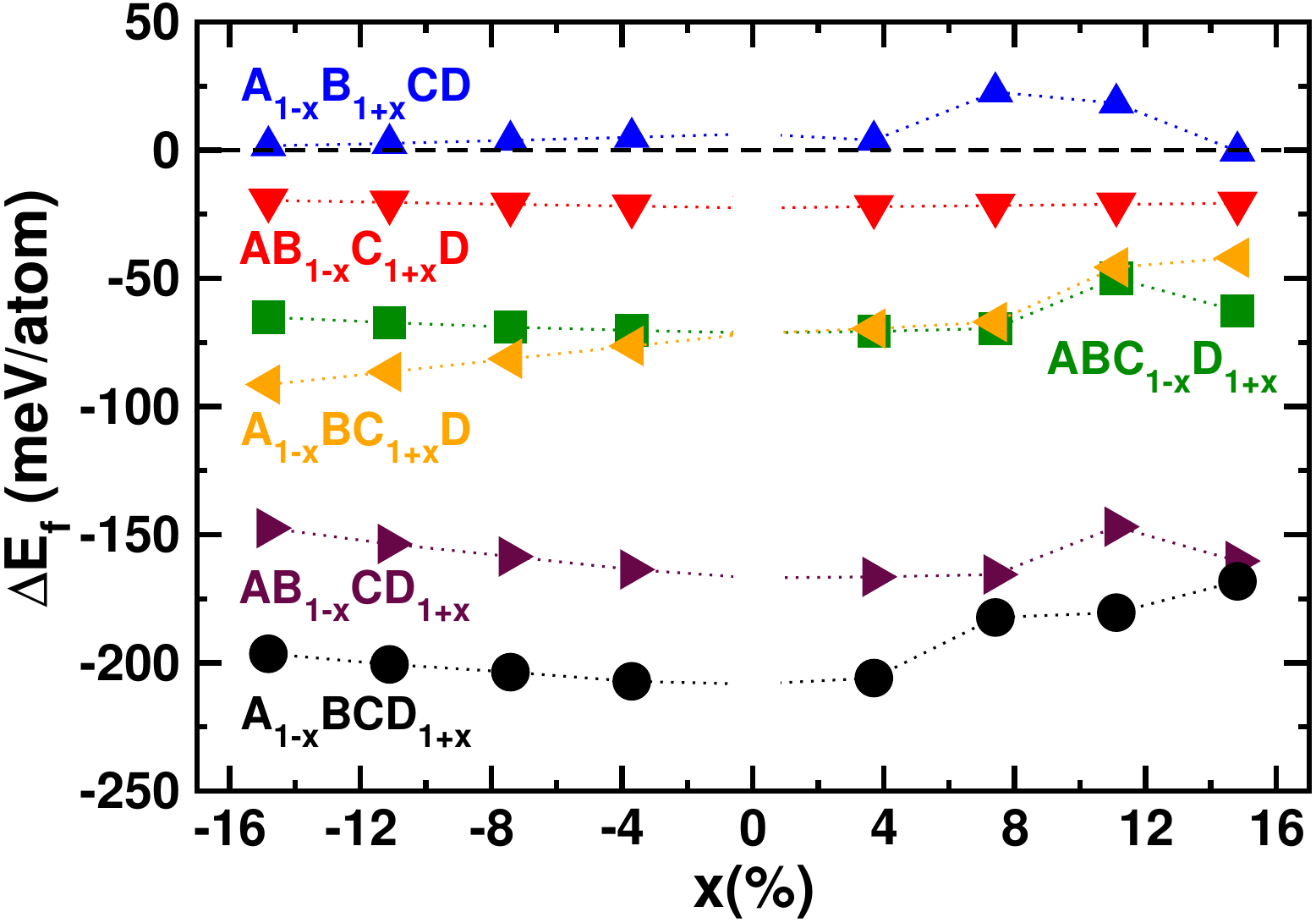}
\caption{Formation energy ($\Delta E_f $) for various binary antisite disorders. A, B, C and D  indicates Co, Mn, Cr and Al, respectively.}
\label{FE-333_CCMA}
\end{figure}

\subsubsection{Point defects (Antisite, Swap, Vacancy)}
Here we investigate the stability and electronic structure of antisite, swap and vacancy defects in CoMnCrAl.  As evidenced in other similar systems,\cite{Raph2,TG,Rav2,Pico1,Pico2} there is a high probability of finding such disorder in these systems. Particularly, antisite disorders up to $14 \%$ are shown to be present in few ternary alloys. We, therefore, have simulated antisite disorder to  $4/27\sim 14.8\%$. We have investigated the stability of all possible combinations of binary antisites as well as  swaps. This should help in predicting the formation of those defects which are most likely to be present in the material.

\begin{figure}[t]
\centering
\includegraphics[scale=0.9]{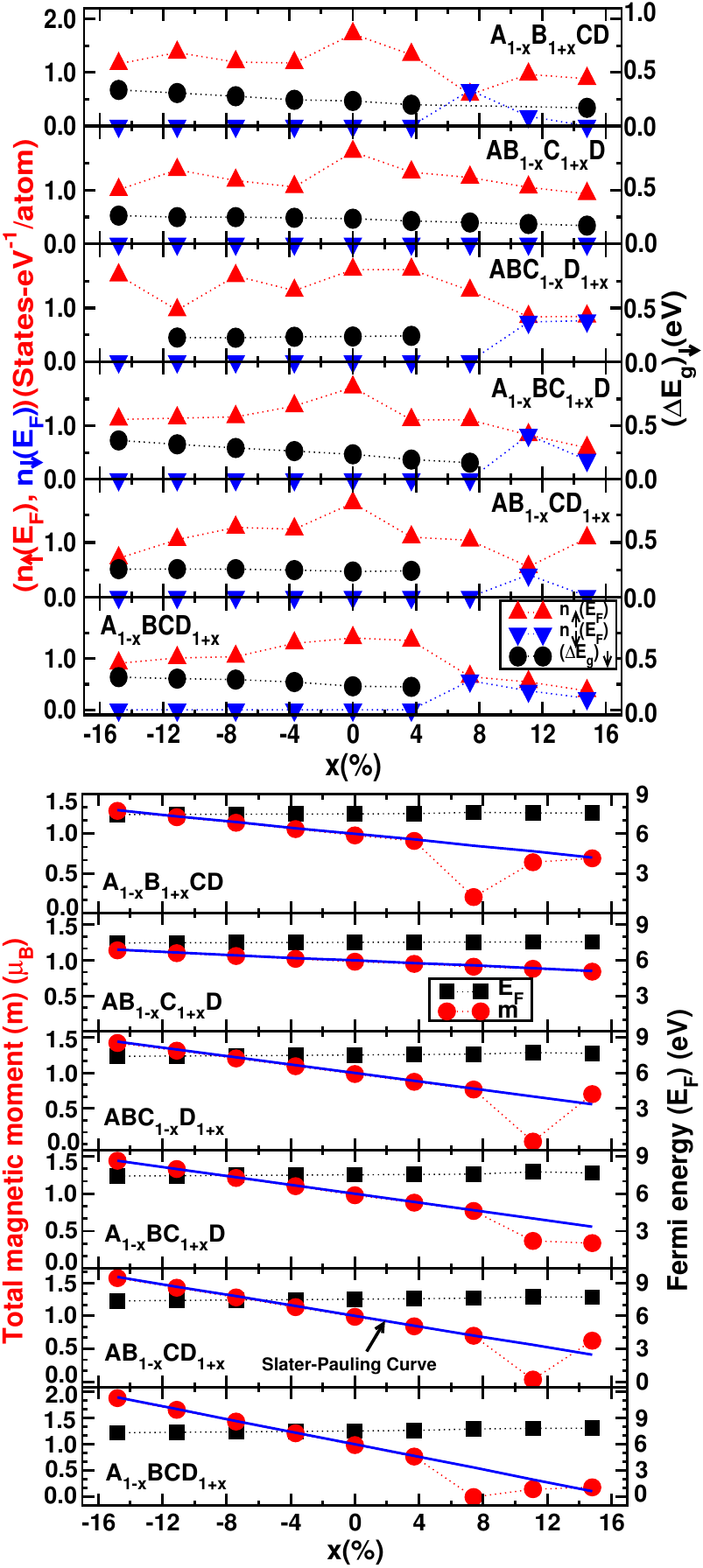}
\caption{For CoMnCrAl, effect of antisite disorder ($x$) on (UP) n$_{\uparrow}(E_\text{F})$ (triangle up), n$_{\downarrow}(E_\text{F})$ (triangle down), and ($\Delta E_\text{g}$)$_{\downarrow}$ (circle) for binary disorders (top to bottom panel); and on (DOWN) total magnetic moments (circle) and $E_\text{F}$ (square). A, B, C and D  indicate Co, Mn, Cr and Al, respectively. The straight lines (down panels) are just a guide to the eye for $m$ vs. $x$ data to check validity of Slater-Pauling rule. }
\label{dos+gap-333_CCMA}
\end{figure}

\begin{figure*}[t]
\centering
\includegraphics[scale=0.8]{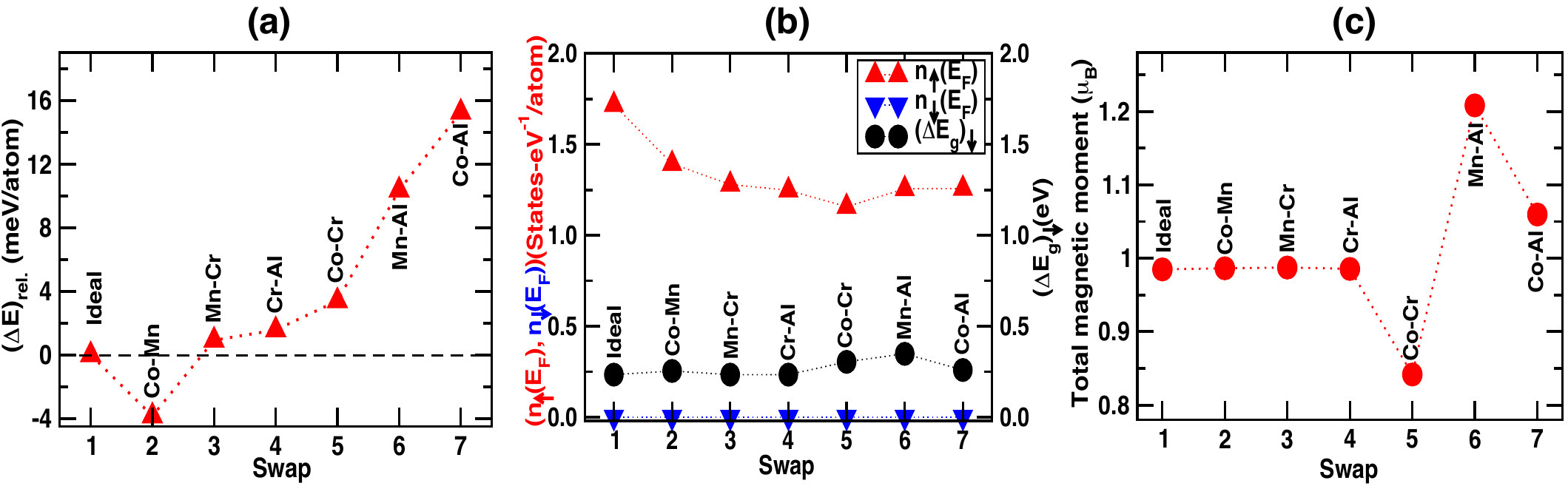}
\caption{For CoMnCrAl, effect of $2$-atom swap on (a) the relative formation energy: ($\Delta E)_{\text{rel.}}$=$(\Delta E_{f})_{swap}$ - $(\Delta E_{f})_{ideal}$; (b) n$_{\uparrow}(E_\text{F})$, n$_{\downarrow}(E_\text{F})$,  ($\Delta E_\text{g}$)$_{\downarrow}$, and (c) total magnetic moments ($\mu_\text{B}$).}
\label{swap_CCMA}
\end{figure*}

\begin{figure}[b]
\centering
\includegraphics[scale=0.5]{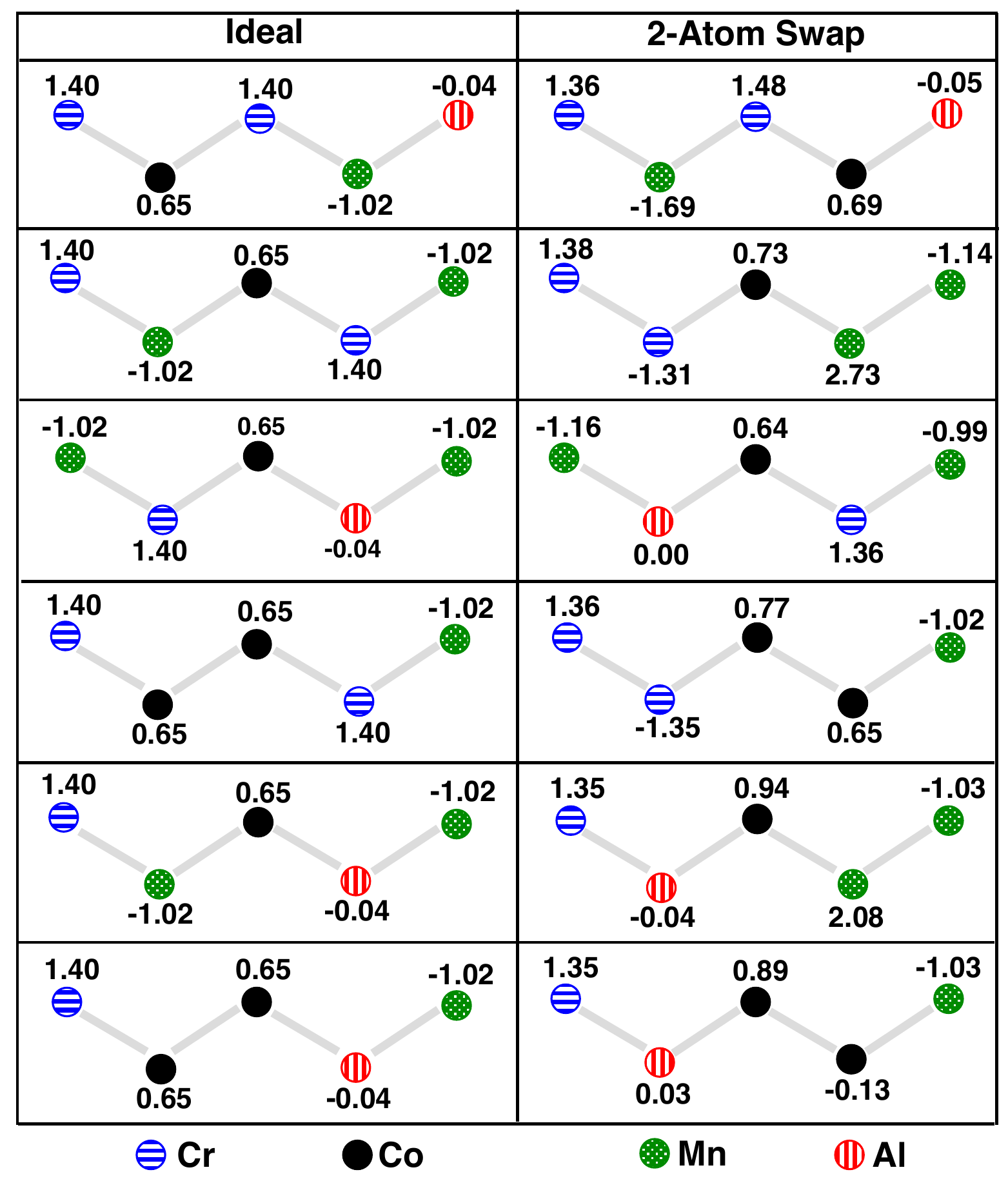}
\caption{For CoMnCrAl, effect of 2-atom swap on moments ($\mu_\text{B}$) at/near swapped sites. Left (Right) column indicates the results without (with) swap.}
\label{swap_pmagmom_CCMA}
\end{figure}

{\it Antisite Defects}:\ \  The $108$-atoms/cell calculations are performed to analyze all binary antisite combinations, i.e., (Co$_{1-x}$Mn$_{1+x}$), (Mn$_{1-x}$Cr$_{1+x}$), (Cr$_{1-x}$Al$_{1+x}$), (Co$_{1-x}$Cr$_{1+x}$), (Mn$_{1-x}$Al$_{1+x}$) and (Co$_{1-x}$Al$_{1+x}$).
Figure \ref{FE-333_CCMA} shows the formation energy ($\Delta E_f$) of CoMnCrAl with  $6$ possible antisites. A, B, C and D indicate the elements Co, Mn, Cr and Al, respectively. +ve (-ve) values of $x$ simply shows excess (deficit) of an element in the system. For example, A$_{1-x}$B$_{1+x}$CD  indicates excess (deficit) of B for +ve $x$ (-ve $x$) over A. It is evident
from the plot that the antisites with respect to (Co,Al) and (Mn,Al) pairs are most likely (lowest formation energy) to be present in the sample, and hence should be observed by neutron diffraction experiment.  The (Co,Mn) antisite has the least probability to occur. Other pairs have intermediate formation energies and may occur during the alloy preparation.

\par For cubic crystals, the condition for mechanical stability among the elastic constants ($C_{ij}$) are
\[
\left(C_{11}-C_{12} \right)/2 > 0, \;\; \left( C_{11}+2C_{12} \right) /3 > 0, \;\; C_{44} > 0. 
\]
This condition is called the ``Born-Huang'' criteria.\cite{BHC}  
The calculated $C_{ij}$  for CoMnCrAl are summarized in Table \ref{tab1}, which clearly satisfies the Born-Huang criteria.
\begin{table}[b]
\begin{ruledtabular}
\caption{Calculated elastic constants C$_{ij}$ (in GPa) for CoMnCrAl system at experimental lattice constant.}
\label{tab1}
\begin{tabular}{c c c c}
   a (\AA) & C$_{11}$ & C$_{12}$ & C$_{44}$ \\
\hline
 5.78 & 222.13 & 95.54 & 93.42\\
\end{tabular}
\end{ruledtabular}
\end{table}

Figure \ref{dos+gap-333_CCMA} shows the DoS at $E_\text{F}$ (majority and minority spin) and minority-spin band gap vs. $x$, the antisite disorder. In each panel,  triangle Up (Down) shows DoS($E_\text{F}$) for spin Up (Down), and solid circle represents the minority-spin band gap ($\Delta E_\text{g}$)$_\downarrow$. One should notice 
different y-scales on the left and right side of vertical axis. Interestingly,
the most stable antisite defect pairs [(Co,Al) and (Mn,Al)] induce a 
transition from half-metallic to metallic state above $\sim 3.7\%$ of 
Al-excess. This is due to a disorder induced state at E$_\text{F}$ in the
minority spin channel which kills the band gap. Similar transitions also
occur in (Co,Cr) and (Cr,Al) antisite pairs which are also likely to form
in CoMnCrAl with relatively lower probability.  (Mn,Cr) antisite pair, on the other
hand, retain the half-metallic character of the alloy throughout the 
concentration ($x$). Another point to notice is a small increase of minority
spin band gap with excess of transition metals over Al. 

Half-metallic to metallic transition, as depicted in Fig. 
\ref{dos+gap-333_CCMA} is intimately connected  to the change in magnetism
of the alloy. Lower panel of Fig. \ref{dos+gap-333_CCMA} shows the variation of total 
magnetic moment and $E_\text{F}$ as a function of $x$. Notably, the total magnetization changes smoothly for all $x$ except the
transition points (half-metallic to metallic) where it takes discontinuous
jump. Such anomalous change in magnetization is not common in Heusler alloy
and will be worth verifying experimentally. The Slater-Pauling (SP) rule\cite{Slat1,Paul1} is a necessary (but still not sufficient) criteria to be satisfied by the Heusler alloy in order to show half-metallic behavior. Most of the antisite binary disordered configurations of the CoMnCrAl alloy follow SP rule (blue solid line of Fig.4 (down)) except few where the magnetic moment changes discontinuously mediated by the phase transition from half-metallic to metallic state. 

Figure \ref{dos+gap-333_CCMA}  shows concentration variation of Fermi energy (E$_{\text{F}}$). The main purpose of showing E$_{\text{F}}$ is to check whether the rigid band behavior holds or not with the introduction of defects. Although such behavior arises quite intuitively in most systems, but there are various exceptions e.g. the antisite defects in Nb$_{1+x}$Fe$_{2-x}$ which gives rise to a non-rigid band behavior.{\cite{Tompsett}} Such behavior occur due to the onset of nearly flat bands in the vicinity of E$_{\text{F}}$ which yields a sharp density of states at E$_{\text{F}}$. This helps in explaining the origin of quantum criticality in this system.  The present case shows no such behavior. 

\begin{figure}[b]
\centering
\includegraphics[scale=0.5]{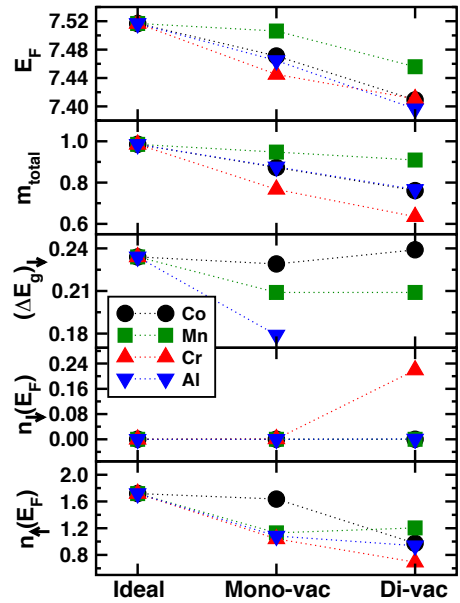}
\caption{For CoMnCrAl, effect of single and double vacancy on $E_{\text{F}}$, total magnetic moments ($m_{\text{total}}$) ($\mu_\text{B}$), ($\Delta E_\text{g}$)$_\downarrow$ (eV), n{$_{\downarrow}$}($E_{\text{F}}$) and n{$_{\uparrow}$}($E_{\text{F}}$) (states-eV$^{-1}$/atom).} 
\label{vacancy_CCMA}
\end{figure}

{\it Swap Antisites}:\ \  Depending on the method of sample preparation, swapping (interchange of 
position between two atoms) is another kind of defect which 
occur in real materials. Such defects (say swap between $A$ and $B$ atoms)
can also be viewed  as the sum of two different $A$ and $B$ atomic antisites
that tend to aggregate. As before, we consider all possible binary swap
among transition metals  as well as the main group elements (Al). Figure 
\ref{swap_CCMA}(a) shows the relative formation energies of different 
combinations of swapping pairs. Energy corresponding to no swap configuration is 
considered as the reference energy $(\Delta E_{f})_{ideal}$ whereas ($\Delta E_{f})_{swap}$ represents formation energy
after swapping. (Co,Mn) swap exhibits the lowest 
formation energy which indicates the possibility of spontaneous formation of
such interchange. Swap involving Al-atom is not favorable. On the other hand
the probability of occurrence of (Mn,Cr) interchange is moderate. DoS at
$E_{\text{F}}$ [$n_{\downarrow}(E_\text{F})$ , $n_{\uparrow}(E_\text{F})$]
and minority spin band gap ($\Delta E_\text{g}$)$_\downarrow$ are shown in 
Fig. \ref{swap_CCMA}(b). Halfmetallicity in CoMnCrAl is quite robust against
swapping with a minor change in the band gap. Majority spin DoS at 
$E_\text{F}$, however, decreases due to the defect induced state.

Interestingly, swaps causes odd  behavior in the total 
magnetization for certain pairs of swapping combinations, e.g., (Co,Cr), 
(Mn,Al), see Fig. \ref{swap_CCMA}(c). Such a behavior 
violates the Slater-Pauling rule in spite of the half-metallic nature of
the alloy. To gain a deeper insight, we have calculated the local moments
at/near the individual atomic sites as shown in Fig. \ref{swap_pmagmom_CCMA}. Left 
panel shows the result for ideal structure (no swap) and the right panel
for swapped structure. In the case of (Co,Cr) swap, Cr (at the swapped site) 
becomes antiferromagnetic with respect to Co unlike their ferromagnetic coupling 
in ideal case, and hence a sharp decrease in total magnetization. In contrast,
Mn becomes ferromagnetic when swapped with Al and forms a large moment 
($\mu_{\text{Mn}}=3.10 \mu_\text{B}$) compared to the ideal case which causes a sharp 
increase in the total magnetic moment. The actual magnetic map of the
 defected structure is somewhat complicated and the effect is found to survive up to the second-nearest neighbors.

\begin{figure}[t]
\centering
\includegraphics[scale=1.0]{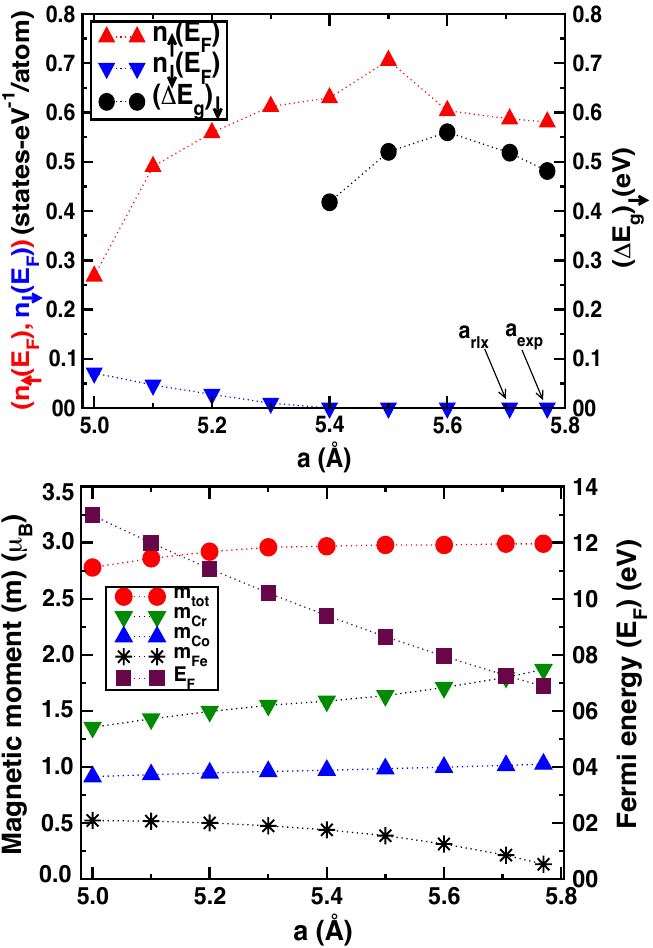}
\caption{Same as Fig. \ref{press_CMCA}, but for CoFeCrGe. Transition occur at $5.4$~\r{A}, with $a_{\text{rlx}}=5.71$ \r{A}
($a_{\text{exp}}=5.77$ \r{A}). }
\label{dos+gap_CCFG}
\end{figure}

{\it Vacancy Defects}:\ \ \  For vacancies created at the three transition-metal sites and the Al sites,  we have checked the effect of both single vacancy ($1$ out of $27 \sim 3.7\% $) and double vacancy ($2$ out of  $27 \sim 7.4\% $). 
Figure \ref{vacancy_CCMA} summarizes the main electronic properties, i.e.,  $E_{\text{F}}$, total moment 
($m_{\text{total}}$), minority-spin band gap ($\Delta E_\text{g}$)$_\downarrow$, 
DoS($E_{\text{F}}$) for spin UP and Down due to the substitution of such 
vacancies. In each panel, circle, square, triangle UP and triangle Down symbols
indicate the results due to vacancies at Co, Mn, Cr and Al positions
respectively. As expected, $E_{\text{F}}$ decreases with the introduction of 
vacancies from rigid band concepts. Due to the reduction in the total number
of valence electrons, Slater Pauling rule may not necessarily hold in all
cases, as shown in the second (from top) panel. A substitution of $3.7\%$ 
($7.4\%$) vacancy at Co, Mn, Cr and Al sites reduces the total number of 
valence electrons of stoichiometric CoMnCrAl by $0.33\ (0.66)$, $0.25\ (0.5)$,
$0.22\ (0.44)$, $0.11\ (0.22)$ respectively. Based on the total moments in 
Fig. \ref{vacancy_CCMA}, none of the vacancy substitutions satisfy the SP rule except for Al. 
Another striking feature is the loss of half-metallicity (zero minority band gap) in case of Cr vacancies. 
Al introduces a small state  at  $E_{\text{F}}$ in the minority-spin DoS 
and makes the system weakly metallic. 
All other vacancies preserve the half-metallicity of the compound.

\subsection{CoFeCrGe}
\subsubsection{Pressure effect}
Figure \ref{dos+gap_CCFG} (top panel) shows the effect of pressure on DoS and band gap (in minority spin channel) for CoFeCrGe.
Unlike the case of CoMnCrAl, half-metallicity in this case is more robust. In other words, 
CoFeCrGe require a much higher pressure ($6-7 \%$ smaller lattice parameter compared to 
$a_{\text{exp}}$) to destroy the half-metallic nature and transit to a metallic state.
Effect of pressure on the magnetic moments are shown in  Fig. \ref{dos+gap_CCFG}. 
Variation in magnetic moment (atom-projected as well 
as total) is very small, indicating, the robustness of ferromagnetic behavior of the
system and hence following the SP rule throughout the pressure range considered
here. Fermi energy indeed gets enhanced under pressure, similar to the case of CoMnCrAl.

\begin{figure}[t]
\centering
\includegraphics[scale=0.48]{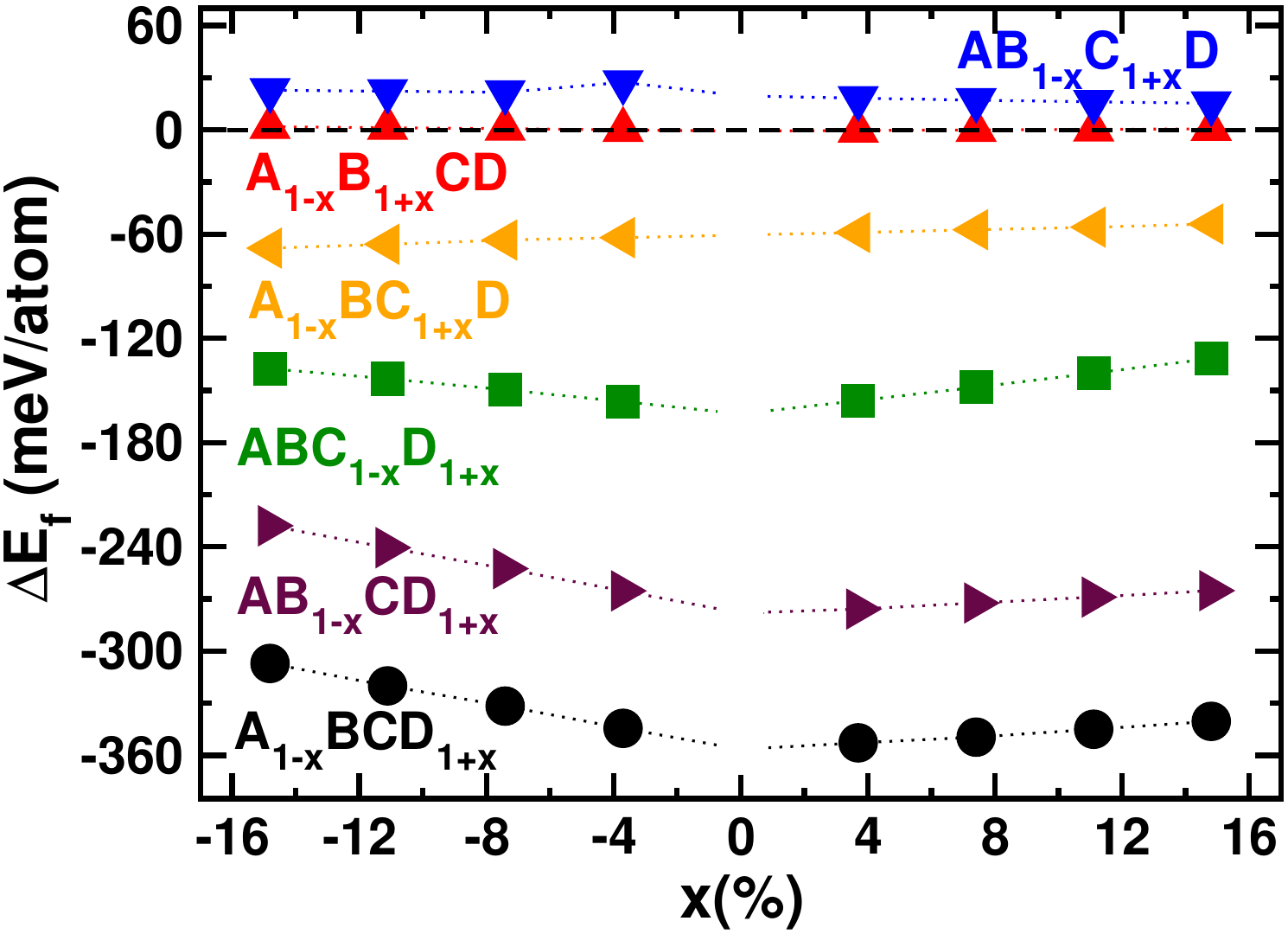}
\caption{Same as Fig. \ref{FE-333_CCMA}, but for CoFeCrGe. Here, A, B, C and D represent Co, Fe, Cr and Ge atom respectively.}
\label{FE-333_CCFG}
\end{figure}

\begin{figure}[t]
\centering
\includegraphics[scale=1.0]{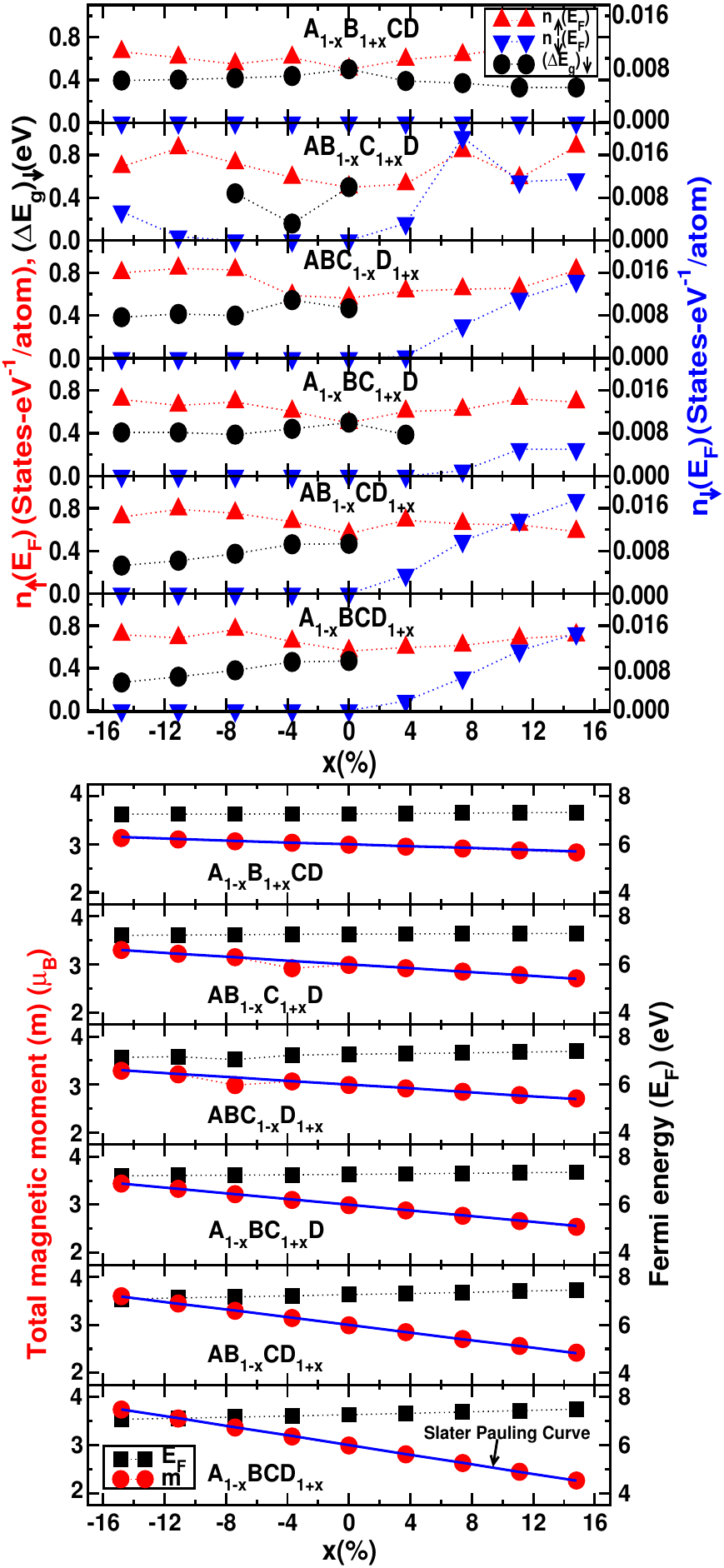}
\caption{Same as Fig. \ref{dos+gap-333_CCMA}, but for CoFeCrGe. A, B, C and D represent Co, Fe, Cr and Ge atom respectively.}
\label{dos+gap-333_CCFG}
\end{figure}

\begin{figure*}[t]
\centering
\includegraphics[scale=0.8]{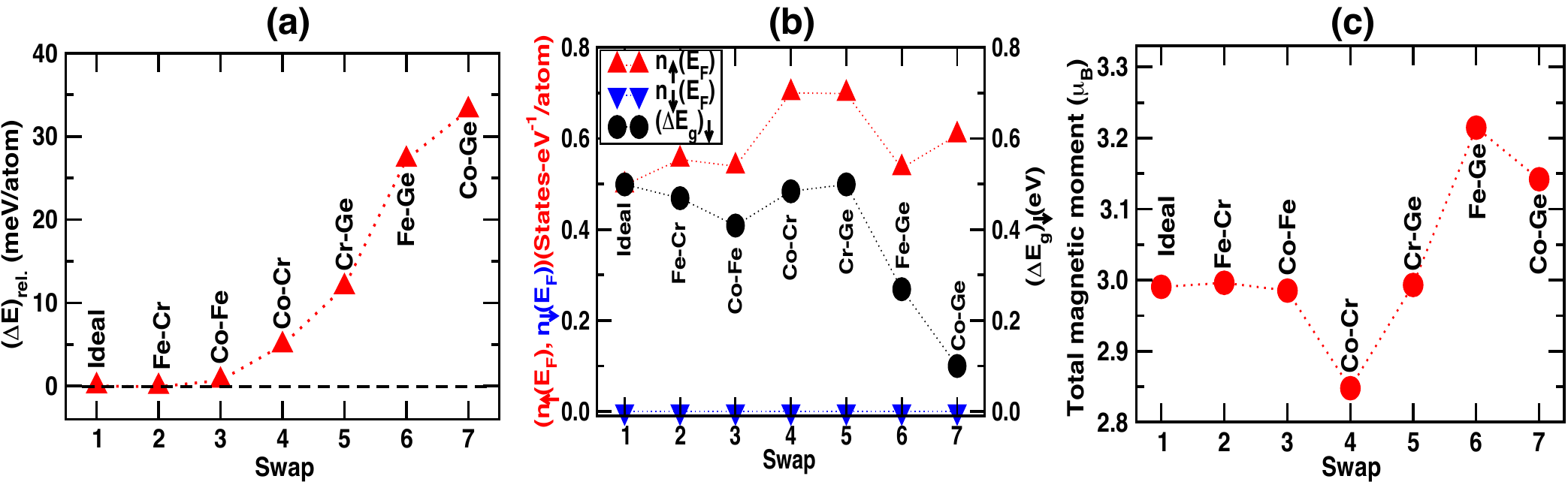}
\caption{Same as Fig. \ref{swap_CCMA}, but for CoFeCrGe.}
\label{swap_CCFG}
\end{figure*}

\subsubsection{Point defects (Antisite, Swap, Vacancy)}

\begin{figure}[b]
\centering
\includegraphics[scale=0.45]{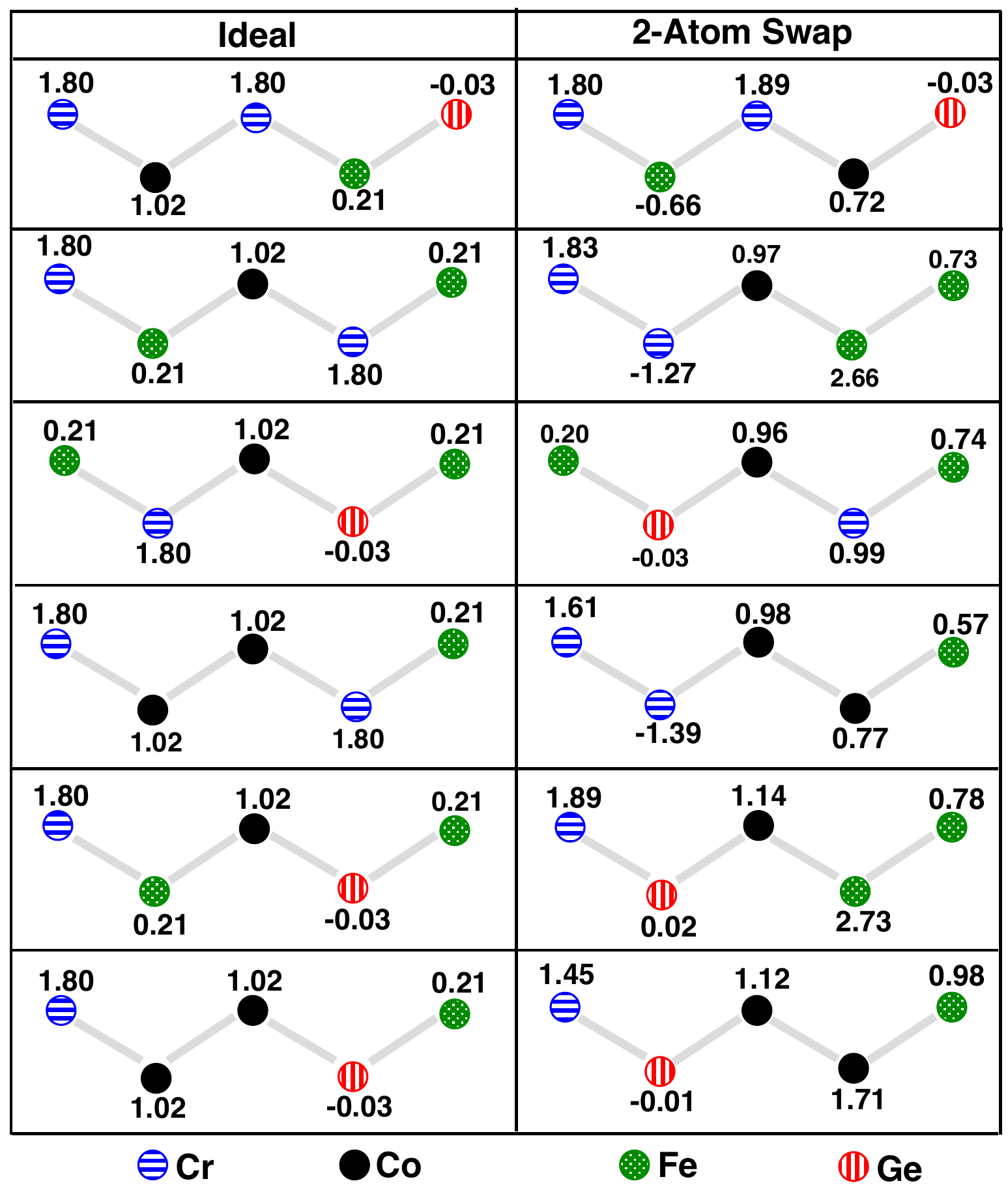}
\caption{Same as Fig. \ref{swap_pmagmom_CCMA}, but for CoFeCrGe.}
\label{swap_pmagmom_CCFG}
\end{figure}

{\it Antisite Defects: } 
All possible combinations of binary antisite disorder are investigated in CoFeCrGe also. 
Formation energies ($\Delta E_f$) for all such antisite pairs in both excess (positive $x$)
and deficit (negative $x$)  range are shown in Fig. \ref{FE-333_CCFG}. (Co,Ge) pair
is found to be energetically the most stable antisite disorder in CoFeCrGe. (Co,Fe) and
(Fe,Cr) pairs are unfavorable to mix and hence are less likely to form during
processing. In terms of energetics, all the pairs including (Co,Ge) in CoFeCrGe are more
stable (from magnitude of $\Delta E_f$) compared to those in CoMnCrAl. In both the alloys we have seen that Co based antisite with the non magnetic element (i.e., Al in CoMnCrAl and Ge in CoFeCrGe) is the most favorable mixing and should be observed experimentally.

\par Elastic constants for CoFeCrGe are given in Table \ref{tab2}. These values are calculated at the experimental lattice constant ($a_{\text{exp}}$). We have also calculated elastic constants at $a=5.72 $ \r{A}, which are similar to those tabulated in the second row of Table \ref{tab2}. It is clear that the Born-Huang criteria is satisfied for CoFeCrGe system also.

\begin{table}[b]
\caption{Calculated elastic constants (in GPa) for CoFeCrGe.}
\label{tab2}
\begin{ruledtabular}
\begin{tabular}{c c c c c}
  & a (\AA) & C$_{11}$ & C$_{12}$ & C$_{44}$ \\
\hline
This work & 5.77 & 207.94 & 184.20 & 108.40\\
Other work\cite{Iyigor} & 5.72 & 193.55 & 192.36 & 120.36\\
\end{tabular}
\end{ruledtabular}
\end{table}

\par Formation energies shown in Fig. \ref{FE-333_CCMA} and \ref{FE-333_CCFG} are with respect to the ternary parent compounds, however one can also estimate the same with respect to the pure elemental ground states. Defect energy is another quantity to investigate the stability of defects. For completeness, we have also calculated these energies for both CoMnCrAl and CoFeCrGe. These results are shown in the supplementary materials.\cite{suppl}

Figure \ref{dos+gap-333_CCFG} (top panel) shows the DoS at $E_{\text{F}}$ (majority and minority spin) and the band gap 
($\Delta E_\text{g}$)$_{\downarrow}$ vs. $x$ for various antisite disorders. One should notice that, unlike CoMnCrAl, the y-scale for n$_{\downarrow}(E_\text{F})$ and ($\Delta E_\text{g}$)$_{\downarrow}$ are interchanged here. This is done to separate the smaller magnitude of n$_{\downarrow}(E_\text{F})$ compared to the large values of n$_{\uparrow}(E_\text{F})$ and ($\Delta E_\text{g}$)$_{\downarrow}$. One of the main differences in CoFeCrGe is the extremely small values of  n$_{\downarrow}(E_\text{F})$ compared to that in CoMnCrAl. A small antisite disorder introduces a very small DoS at $E_{\text{F}}$
in minority spin channel in selected x-range, causing a transition
from half-metallic to metallic state. Unlike CoMnCrAl, the magnitude of n$_{\downarrow}(E_\text{F})$ is so small that it may be difficult to gauge whether a transition will really happen in a real sample. As such, we expect the half-metallic to metallic transition to be more robust in CoMnCrAl  than CoFeCrGe. Variation of magnetic moment ($m$) vs. $x$ in the present case is relatively more monotonous compared to that in CoMnCrAl, which may be attributed to a much smaller jump of n$_{\downarrow}(E_\text{F})$ at the transition point. Another key difference is the robustness of ferromagnetic behavior throughout the concentration $x$ in the present case unlike CoMnCrAl where the half-metallic to metallic transition is often mediated by a magnetic transition (antiferromagnetic to magnetic).

{\it Swap Antisites: } \ 
Effect of interchanging the position of one atom (in a 3 $\times$ 3 $\times$ 3 supercell) on the electronic and magnetic properties of CoFeCrGe is shown in Fig. \ref{swap_CCFG}.  Positive relative formation energies ($\Delta E)_{\text{rel}}$ suggest swapping of atoms to be quite unlikely during the formation. (Co,Fe) and (Fe,Cr) swapping may have a very small probability to occur. Although the band gap ($\Delta E_\text{g}$)$_{\downarrow}$ changes dramatically for some swapping pairs, half-metallicity is preserved in all cases. This is similar to the case of CoMnCrAl. (Fe,Cr) and (Cr,Ge) swaps almost give the same total magnetic moments as the ideal (no swap) case (Fig. \ref{swap_CCFG}c). On the other hand, (Co,Cr), (Fe,Ge) and (Co,Ge) pairs have the strongest effect on the total moment. In a 3 $\times$ 3 $\times$ 3 supercell, the total moment decreases by $3.84 \mu_\text{B}$ for (Co-Cr) swap and increases by $6.06 \mu_\text{B}$ and $4.09 \mu_\text{B}$ for (Fe,Ge) and (Co,Ge) swap respectively as compared to the ideal case.

Individual magnetic moments at/near the defective sites show quite interesting behavior as depicted in Fig \ref{swap_pmagmom_CCFG}. Even though, swapping between Fe-Cr, Co-Fe and Cr-Ge gives almost the same value of total magnetic moment as in the ideal case (Fig. \ref{swap_CCFG}c), the magnitude of individual magnetic moments at the defective sites are somewhat different in each case  with respect to equivalent sites of ideal structure. (Fe,Cr) swapping pair is quite interesting out of the three, where Cr becomes antiferromagnetic (m$_{\text{Cr}}$ changes from $1.8 \mu_\text{B}$ in ideal case to $-1.27 \mu_\text{B}$) and Fe gains a huge moment (m$_{\text{Fe}}$ goes from $0.21 \mu_\text{B}$ to $2.66 \mu_\text{B}$) after swapping. In addition to the swapped sites, magnetic moments of the neighboring sites (nearest and next nearest neighbors) are also affected which collectively sum up to yield similar total moment as the ideal case. Swapping of Co with Cr causes an antiferromagnetic alignment of Cr along with a reduction of Co moment (keeping its ferromagnetic nature intact) resulting in an overall reduction of total moment. However in the case of (Fe,Ge) and (Co,Ge) swaps, both Fe and Co gain a moment of $2.52 \mu_\text{B}$ and $0.7 \mu_\text{B}$ respectively, resulting in an overall enhancement of the total moment of the cell. In all the swapping cases, the magnetic interactions do not only affect the nearest neighbors of the swapped sites but also the next nearest neighbors beyond which the effect becomes negligibly small.

\begin{figure}[b]
\centering
\includegraphics[scale=0.5]{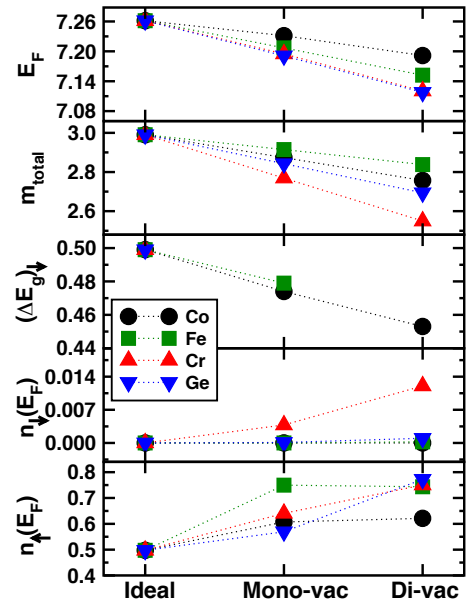}
\caption{Same as Fig. \ref{vacancy_CCMA}, but for CoFeCrGe. }
\label{vacancy_CCFG}
\end{figure}

{\it Vacancy Defects: }\ \
 Single and double vacancy effects on electronic and magnetic properties of CoFeCrGe are shown in Fig. \ref{vacancy_CCFG}. 
In each panel, circle, square, triangle up and triangle down symbols indicate the results due to vacancies at Co, Fe, Cr and Ge positions respectively. Vacancies reduce E$_{\text{F}}$ as expected from rigid band concept. Magnetization is least (most) affected by Fe (Cr) vacancies which is due to the lowest (highest) magnetic moment of Fe (Cr) atoms in the compound. Interestingly, Ge vacancies cause a reduction of the moments of its neighboring atoms and hence an effective reduction of the total cell moment. Vacancies at Cr and Ge sites result in a half-metallic to metallic transition with a very small disorder induced DoS (at $E_{\text{F}}$) in the minority spin channel. Magnitude of the vacancy induced state (in minority channel) in the present case is extremely small as compared to that in CoMnCrAl.

\newpage
\section{Conclusions}
We  have performed detailed first-principles calculations on two quaternary Heusler alloys, CoCrMnAl and CoCrFeGe, to determine the effect of hydrostatic pressure and various intrinsic defects (antisite disorder, pairwise swap, and vacancies) on their electronic and magnetic properties, as well as to assess the most favorable defects. These two systems are interesting because of their high $T_\text{C}$ and partial availability of experimental data. Understanding the effects of operative defects provide unique tool to control and develop the best materials for spintronics based applications. 

\par We find antiferromagnetic (ferromagnetic) alignment of Cr-moments with respect to other transition elements in CoMnCrAl (CoFeCrGe). CoMnCrAl is found to be quite sensitive to pressure, and undergoes a half-metallic--to--metallic transition with a $2-3 \%$ reduction in lattice parameter; CoFeCrGe properties are much more robust against pressure.  Also, in contrast to CoFeCrGe, CoMnCrAl is quite sensitive to these defects. Formation energies provide details on stability of the defects, as well as which order they can form during processing a real sample. Above a certain antisite defect concentration, CoMnCrAl undergoes a half-metallic--to--metallic transition mediated by a concomitant magnetic transition. Halfmetallicity is quite robust against swap defects in both systems. Magnetization (both local and bulk) is found to behave sensitively with swapping from their preferred Wyckoff positions, often changing spin orientation of the antisite atoms -- violating that expected from the Slater-Pauling rule. Vacancies are found to cause narrowing of the band gap as compared to the ideal structure, including vanishing (a transition) in some cases. Clearly, it is crucial to prevent thermal-induced or growth defects during the synthesis of these alloys, which are used as electrodes in the magneto electronic devices for spintronic applications.

\section*{Acknowledgement}
Enamullah (an institute post-doctoral fellow) acknowledges IIT Bombay for funding to support this research.
Work at Ames Lab was supported by the U.S. Department of Energy (DOE), Office of Science, Basic Energy Sciences, Materials Science and Engineering Division. Ames Laboratory is operated for the U.S. DOE by Iowa State University under contract \#DE-AC02-07CH11358.




\begin{thebibliography}{99}
\bibitem{Raph1} M. P. Raphael, B. Ravel, M. A. Willard, S. F. Cheng, B. N. Das, R. M. Stroud, K. M. Bussmann, J. H. Claassen, and V. G. Harris, Appl. Phys. Lett. {\bf{79}}, 4396 (2001).
\bibitem{Rav1} B. Ravel, J. O. Cross, M. P. Raphael, V. G. Harris, R. Ramesh, and L. V. Saraf, Appl. Phys. Lett. {\bf{81}}, 2812 (2002).
\bibitem{Raph2} M. P. Raphael, B. Ravel, Q. Huang, M. A. Willard, S. F. Cheng, B. N. Das, R. M. Stroud, K. M. Bussmann, J. H. Claassen, and V. G. Harris, Phys. Rev. B {\bf{66}}, 104429 (2002).
\bibitem{Groot} R. A. de Groot, F. M. Mueller, P. G. van Engen, and K. H. J. Buschow, Phys. Rev. Lett. {\bf{50}}, 2024 (1983).
\bibitem{Zhu} Z. H. Zhu, and X. H. Yan, J. Appl. Phys. {\bf{106}}, 023713 (2009).
\bibitem{Kob} K. -I. Kobayashi, T. Kimura, H. Sawada, K. Terakura, and Y. Tokura, Nature {\bf{395}}, 677 (1998).
\bibitem{Dho} J. Dho, S. Ki, A. F. Gubkin, J. M. S. Park, and E. A. Sherstobitova, Solid State Commun. {\bf{150}}, 86 (2010).
\bibitem{Soe} S. Soeya, J. Hayakawa, H. Takahashi, K. Ito, C. Yamamoto, A. Kida, H. Asano, and M. Matsui, Appl. Phys. Lett. {\bf{80}}, 823 (2002).
\bibitem{Nou} A. Nourmohammadi, and M. R. Abolhasani, Solid State Commun. {\bf{150}}, 1501 (2010).
\bibitem{Wan} W. Z. Wang, and X. P. Wei, Comput. Mater. Sci. {\bf{50}}, 2253 (2011).
\bibitem{Kro} L. Kronik, M. Jain, and J. R. Chelikowsky, Phys. Rev. B {\bf{66}}, 041203(R) (2002).
\bibitem{Noo} N. A. Noor, S. Ali, and A. Shaukat, J. Phys. Chem. Solids {\bf{72}}, 836 (2011).
\bibitem{Far} R. Farshchi, and M. Ramsteiner, J. Appl. Phys. {\bf{113}}, 191101 (2013).
\bibitem{Has1} M. Hashimoto, J. Herfort, H. -P. Schonherr, and K. H. Ploog, Appl. Phys. Lett. {\bf{87}}, 102506 (2005).
\bibitem{Has2} M. Hashimoto, A. Trampert, J. Herfort, and K. H. Ploog, J. Vac. Sci. Technol. B {\bf{25}}, 1453 (2007).
\bibitem{Has3} M. Hashimoto, J. Herfort, A. Trampert, H. -P. Schonherr, and K. H. Ploog, J. Phys. D: Appl. Phys. {\bf{40}}, 1631 (2007).
\bibitem{Ozd} K. Ozdogan, E. Sasioglu, and I. Galanakis, J. Appl. Phys. {\bf{113}}, 193903 (2013).
\bibitem{Ali} V. Alijani, J. Winterlik, G. H. Fecher, S. S. Naghavi, and C. Felser, Phys. Rev. B {\bf{83}}, 184428 (2011).
\bibitem{Dai} X. Dai, G. Liu, G. H. Fecher, C. Felser, Y. Li, and H. Liu, Appl. Phys. Lett. {\bf{105}}, 07E901 (2009).
\bibitem{Sin} M. Singh, H. S. Saini, and M. K. Kashyap, J. Mater. Sci. {\bf{48}}, 1837 (2013).
\bibitem{Ali1} V. Alijani, J. Winterlik, G. H. Fecher, S. S. Naghavi, S. Chadov, T. Gruhn, and C. Felser, J. Phys: Condens. Matter {\bf{24}}, 046001 (2012).
\bibitem{Gal} I. Galanakis, K. Ozdogan, E. Sasioglu, and B. Aktas, Phys. Rev. B {\bf{75}}, 172405 (2007).
\bibitem{Luo} H. Z. Luo, H. W. Zhang, Z. Y. Zhu, L. Ma, S. F. Xu, G. H. Wu, X. X. Zhu, C. B. Jiang, and H. B. Xu, J. Appl. Phys. {\bf{103}}, 083908 (2008).
\bibitem{Gal1} I. Galanakis, and E. Sasioglu, Appl. Phys. Lett. {\bf{99}}, 052509 (2011).
\bibitem{Mei} M. Meinert, Jan-Michael Schmalhorst, C. Klewe, G. Reiss, E. Arenholz, T. Bohnert, and K. Nielsch, Phys. Rev. B {\bf{84}}, 132405 (2011).
\bibitem{Rav2} B. Ravel, M. P. Raphael, V. G. Harris, and Q. Huang, Phys. Rev. B {\bf{65}}, 184431 (2002).
\bibitem{TG} T. Graf, C. Felser, and S. S. P. Parkin, Progress in Solid State Chemistry {\bf{39}}, 1 (2011).
\bibitem{Pico1} S. Picozzi, A. Continenza, and A. J. Freeman, J. Appl. Phys.  {\bf{94}}, 4723 (2003).
\bibitem{Pico2} S. Picozzi, A. Continenza, and A. J. Freeman, Phys. Rev. B  {\bf{69}}, 094423 (2004).
\bibitem{Hamad11} B. Hamad, and Qing-Miao Hu, Phys. Satus Solidi B  {\bf{248}}, 2893 (2011).
\bibitem{enam} Enamullah, Y. Venkateswara, S. Gupta, M. R. Varma, P. Singh, K. G. Suresh, and A. Alam, Phys. Rev. B  {\bf{92}}, 224413 (2015).
\bibitem{VASP} G. Kresse, and J. Furthmuller, Phys. Rev. B  {\bf{54}}, 11169 (1996); Comput. Mater. Sci. {\bf{6}}, 15 (1996).
\bibitem{PAW} G. Kresse, and D. Joubert, Phys. Rev. B  {\bf{59}}, 1758 (1999).
\bibitem{BHC} M. Born and K. Huang, {\it{Dynamical Theory of Crystal Lattices}} (Oxford University Press, Oxford, UK, 1954).
\bibitem{Slat1} J. C. Slater, Phys. Rev. {\bf{49}}, 931 (1936).
\bibitem{Paul1} L. Pauling, Phys. Rev. {\bf{54}}, 899 (1938).
\bibitem{Tompsett} Aftab Alam and D. D. Johnson, \PRL {\bf 107}, 206401 (2011).
\bibitem{suppl} See supplementary material at [URL] for details on defect energies and relative formation energies.
\bibitem{Iyigor} A. Iyigor and S. Ugur, Phil. Maga. Lett. {\bf 94(11)}, 708 (2014).
\end{thebibliography}
\end{document}